\documentclass[twocolumn,showpacs,showkeys,
  preprintnumbers,amsmath,amssymb,page]{revtex4}
\usepackage{graphicx}
\usepackage{dcolumn}
\usepackage{bm}

\begin{document}

\preprint{}

\title{Altruistic Duality in Evolutionary Game Theory}

\author{Taksu Cheon}
\email[E-mail:]{taksu.cheon@kochi-tech.ac.jp}
\homepage[\\]{http://www.mech.kochi-tech.ac.jp/cheon/}
\affiliation{
Laboratory of Physics,
Kochi University of Technology,
Tosa Yamada, Kochi 782-8502, Japan
}

\date{August 28, 2003}

\begin{abstract}
A game-theoretic dynamical model of 
social preference and enlightened self-interest
is formulated.
Existence of symmetry and duality in the game matrices with altruistic
social preference is revealed.  The model quantitatively describes 
the dynamical evolution of altruism in prisoner's dilemma and 
the regime change in prey-predator dynamics.
\end{abstract}

\maketitle

%
%
%
Through the modeling of ecosystems, 
evolutionary game theory brings such diverse fields as 
biology, ecology, economics and sociology 
under the umbrella of mathematical 
sciences \cite{MA74,MS82,GI00}.
One central objective of evolutionary game theory is 
to understand the workings of cooperative behavior 
among the individuals in an ecosystem.
Since the publication of the work by Axelrot 
on the prisoner's dilemma \cite{AX84}, it is generally understood
that the concept of altruism holds a key to the
emergence of cooperative behavior.
While egoism, defined as the drive toward maximization 
of individual payoffs, is a cornerstone of game theory, 
casual observation reveals that altruism is just as universal a feature
as egoism in systems consisting of like individuals.
Altruism has obtained mathematical expression
in the work of Bester and G{\"u}th \cite{BG98}, in which 
altruistic behavior is shown to become evolutionarily stable 
in certain situations through the enhancement of 
fitness of a majority of individuals in a system. 
While their mathematical treatment is general and elegant, 
it is formulated in static and descriptive language.
Bringing in {\em dynamics} to the model
would give it more predictive power.  
%
%

In this article, we do not try to explain the emergence of altruistic
cooperation.  
Rather,
we intend to develop a game-theoretic model
of ecosystem whose evolution is driven by the development
of an optimal degree of altruism.  
Toward this goal,
the separation of two time
scales, one for the fast variation of dynamical variables, 
and the other for the secular variation of ``environmental''
coefficients, proves to be crucial \cite{CH03}.
The formulation of our model in terms
of parametric game matrices reveals symmetry properties
of altruistic game theory. 
We demonstrate the usefulness of our approach
through numerical analyses of a game of prisoner's dilemma and a
prey-predator system.
In the latter example, we point out the existence of 
a regime change phenomenon
caused by dynamical symmetry breaking.
%

%
%
We start by considering a system 
of $N$ identical individuals 
randomly paired to repeatedly play
a two-player game with $M+1$ pure strategies. 
We introduce  the payoff matrix $A$ and 
the system average strategy vector ${\bf x}$,
\begin{eqnarray}
\label{defax}
A &=& \{A_{ij}\}
\quad (i,j = 0, ..., M),
\\ \nonumber
{\bf x} &=& \{x_{i}\}
\quad (i = 0, ..., M).
\end{eqnarray}
All entries in $A$ and ${\bf x}$ are real numbers, and
the relation $x_0 = 1 - x_1 - ... - x_M$ is imposed.
It is convenient to consider ${\bf x}$ as column vector so that the matrix
product $A{\bf x}$ is again a column vector.
We index the elements of $A$ such that a player with $i$-th strategy
playing against another player with $j$-th strategy will
obtain the payoff $A_{i,j}$. 
We interpret ${\bf x}$ either as the system being made up of
$x_i N$ players playing $i$-th strategy, or alternatively as
$N$ individuals adopting identical
mixed strategies in which the probability of playing $i$-th strategy is
given by $x_i$.
A player with a mixed
strategy specified by a vector ${\bf s}$ in the system obtains the payoff
%
\begin{eqnarray}
\label{matx}
\left<{\bf s} | A | {\bf x} \right>
= {\bf s}^\dagger A {\bf x}
= {\bf s}^\dagger {\bf p}(A, {\bf x}).
\end{eqnarray}
In the second equality,  
the payoff vector is defined as ${\bf p}(A, {\bf x})$$\equiv$$A{\bf x}$, whose
$i$-th element represents 
the payoff of a player with $i$-th pure strategy.
We can average $\left<{\bf s} | A | {\bf x} \right>$ over 
the entire system by the identification ${\bf s}$ $={\bf x}$,  and obtain
the average per capita payoff of the system
$\Pi({\bf x}) \equiv \left<{\bf x} | A | {\bf x} \right> $.
In spite of the use of
bra-ket notations $\left<{\bf s} \right|$ and $\left| {\bf x} \right>$,
an obvious adoption from the quantum mechanics, all entries to the vectors 
are real numbers representing probabilities, and there should be no confusion
on the fact that we are dealing with {\em classical} game theory.
%
%
%

As is well known,
the best strategy for a game among 
players seeking immediate individual payoff maximization
is given by the mixed Nash equilibrium \cite{NA50} of the matrix $A$.
This equilibrium, however, does not always give
maximization of the system average payoff.  
%
%
A system consisting of players with longer view on their payoff
often has a higher average payoff
than a Hobbsian system consisting of narrowly egoistic players. 
To describe such ``enlightened self-interest'', we follow 
Bester and G{\" u}th \cite{BG98}, and separate the process
of reproduction from that of selection:  The players switch
strategies in  pursuit of
the maximization of a range of {\em perceived} payoffs which are related, 
but not necessarily
identical to the real payoff.  
The deviation of perceived from real payoffs could represent 
imperfect information, socially imposed norm, or just error and caprice.
The system is then assumed to be under
slow selection process during which the players with inferior 
(real) payoff are pruned off.
To formulate such two-stage evolution in a simple manner, 
we define a one-parameter  family of 
matrix $A^\kappa$, which we call the {\em game matrix}
with the {\em preference parameter} $\kappa$.
We assume that  $A^\kappa$ reduces to the original payoff
matrix $A$ with $\kappa = 0$.
The specific example we study is the game matrix
%
\begin{eqnarray}
\label{akap}
A^\kappa \equiv (1-\kappa)A+\kappa A^\dagger,
\quad
\kappa\in[0,1] ,
\end{eqnarray}
where $A^\dagger$ is the transposed matrix of $A$
defined by
%
\begin{eqnarray}
\label{adag}
A^\dagger = \{ A^\dagger_{ij} \}
\quad (i,j = 0, ..., M); 
\quad
A^\dagger_{ij} \equiv A_{ji}.
\end{eqnarray}
Notice that, in all instances in this article, superscript $\kappa$ on 
game matrix $A$ and strategy vector ${\bf x}$ 
signify the preference parameter, {\em not} the exponentiation.
Be prepared to see the notations $A^0 = A$, $A^1=A^\dagger$ {etc}..
The meaning of $A^\dagger$ becomes evident by considering
its payoff vector ${\bf p}(A\dagger, {\bf x})$ $= A^\dagger {\bf x}$, 
whose $i$-th component is the average payoff yielded to the
opponent by a player with the pure strategy $i$.  Therefore, we call
$A^\dagger$ the {\em altruistic dual matrix} of the original
payoff matrix $A$.  Obvously we have a relation  
$\left<{\bf x} | A | {\bf x} \right>$
=$\left<{\bf x} | A^\dagger | {\bf x} \right>$,
which simply means that per capita system payoff
can be calculated from the payoff obtained by players,
or from the payoff contributed by him to the rest of the system. 
We can generalize this result for $A^\kappa$ as 
 ${A^\kappa}^\dagger$ $= A^{1-\kappa}$, namely
$A^{1-\kappa}$ is the altruistic dual matrix of $A^\kappa$.
Per capita system payoff $\Pi({\bf x})$
can be calculated from $A^\kappa$ with 
any allowed value of $\kappa$;
%
\begin{eqnarray}
\label{egal}
\Pi({\bf x})
=\left<{\bf x} | A^\kappa | {\bf x} \right> ,
\quad
\kappa\in[0,1].
\end{eqnarray}
%

%
%
Consider a system with a given 
preference parameter $\kappa$ which 
evolves with {\em replicator dynamics} \cite {HS88}, in which
a player with strategy ${\bf s}$ tinkers with the changes
to the strategy by $\delta{\bf s}$, and accepts the changes
in a probability proportional to  the resulting gain in the
payoff $\left<{\bf s}+\delta{\bf s} | A^\kappa | {\bf x} \right>$
$-\left<{\bf s} | A^\kappa | {\bf x} \right>$.
The time development of strategy ${\bf x^\kappa}$ is 
described by the $M$-dimensional Lotka-Volterra equation
%
%
%
%
\begin{eqnarray}
\label{evol}
{ {\dot {x}^\kappa_i} \over {x^\kappa_i} } 
&=& 
\left.
  \partial_{s_i} \!
  \left<{\bf s}|A^\kappa|{\bf x}^\kappa\right> 
\right|_{{\bf s}={\bf x}^\kappa}
\\ \nonumber
&=& {p_i}(A^\kappa,{\bf x}^\kappa)-{p_0}(A^\kappa, {\bf x}^\kappa) ,
\end{eqnarray}
with $i$ $=1, ..., M$.
We refer to this dynamics as an $A^\kappa$-constrained game, 
or simply an $A^\kappa$-game.
Typically, after some time period, ${\bf x}^\kappa$ approaches
a stable fixed point  ${\bf X}^\kappa$,
which we call $A^\kappa$-Nash equilibrium,
that is obtained by the linear equation
%
\begin{eqnarray}
\label{nashkap}
p_i(A^\kappa,{\bf X}^\kappa)=p_0(A^\kappa, {\bf X}^\kappa),
\quad (i=1,...,M) .
\end{eqnarray}
Suppose we have an {\it ensemble} of systems with various
values of preference parameter $\kappa$.  If there is a selection
process based on the payoff $\Pi({\bf X}^{\kappa})$ at work, 
the average preference parameter $\kappa$ shall
evolve toward $\kappa = \kappa_{max}$
that gives the maximum per capita system payoff
$\Pi(X^{\kappa_{max}})$.  For example, we can postulate
\begin{eqnarray}
\label{evolkap}
m {{\dot {\kappa}} \over {\kappa}} 
= 
\partial_{\kappa}
\left<{\bf X}^\kappa|A|{\bf X}^\kappa\right>,
\end{eqnarray}
where $m$ is a large number $m \gg1$ that ensures slow secular variation
of $\kappa$ in comparison to the  variation of dynamical 
variable ${\bf x}^\kappa$. 
We might alternatively consider the development of $\kappa$ by
Newtonian dynamics, in which case $-\Pi({\bf X}^\kappa)$ 
should be identified as the potential. 

Our task is reduced to evaluating the functional profile 
of $\Pi({\bf X}^{\kappa})$.
Let us note the relation
$\Pi({\bf X}^\kappa) 
= p_i(A^\kappa, {\bf X}^\kappa)$ for arbitrary $i$,
which is obtained from (\ref{egal}) and
 (\ref{nashkap}).
Combining this with another equality
%
\begin{eqnarray}
\label{dualprf}
p_k(A^\kappa, {\bf X}^\kappa)
&=&
\sum_{i}{ X_i^{1-\kappa} 
                \sum_{j}A^\kappa_{ij} 
                                      X_j^\kappa }
\\ \nonumber
&=&
\sum_{i}{ X_i^\kappa} 
                \sum_{j}{A^{1-\kappa}_{ij} 
                                      X_j^{1-\kappa} } 
\\ \nonumber
&=& p_{l}(A^{1-\kappa}, {\bf X}^{1-\kappa}) ,
\end{eqnarray}
which is valid for arbitrary $k$ and $l$, 
we obtain the {\em altruistic duality}
%
\begin{eqnarray}
\label{dualityk}
\Pi({\bf X}^\kappa)
=\Pi({\bf X}^{1-\kappa}) .
\end{eqnarray}
Namely, the per capita system payoff for
an $A^\kappa$-game  is exactly equal to 
that of its dual game $A^{1-\kappa}$.
Specifically, we have $\Pi({\bf X}^0)$$= \Pi({\bf X}^1)$, 
an equivalence of payoff between
a completely egoistic game $A$ and a completely
altruistic game $A^\dagger$.
We stress that this duality is nontrivial, unlike, for example,
the mere matrix symmetry (\ref{egal}).
One immediate result of (\ref{dualityk}) is that $\kappa=1/2$
has to be an extremum of the payoff $\Pi({\bf X}^\kappa)$.
If this is the sole maximum, 
we have the inequality
%
\begin{eqnarray}
\label{eqmx}
\Pi({\bf X}^{1/2}) \geqslant \Pi({\bf X}^{\kappa})
\geqslant \Pi({\bf X}^0) .
\end{eqnarray}
In general, there could be other extrema and also $\kappa = 1/2$ could
be a minimum.  But we shall show in the following examples, that there
are indeed interesting cases in which (\ref{eqmx}) holds, and that examples
include the classic prisoner's dilemma.
The {\em maximum happiness of the maximum majority} is achieved
when every individual in the system is constrained to pursue an
equal mixing of egoistic and altruistic payoff.
In hindsight, this is to be expected, since direct maximization of 
the per capita system payoff results in symmetrization
of the game matrix $A$;
%
\begin{eqnarray}
\label{direct}
  \partial_{x_i} \!
  \left<{\bf x}|A|{\bf x}\right>
=
 \left.
   \partial_{s_i} \!
   \left<{\bf s}|(A+A^\dagger)|{\bf x}\right>
 \right|_{{\bf s}={\bf x}}
= 0 .
\end{eqnarray}
%
%

%
%
\begin{figure}
\includegraphics[width=6.0cm]{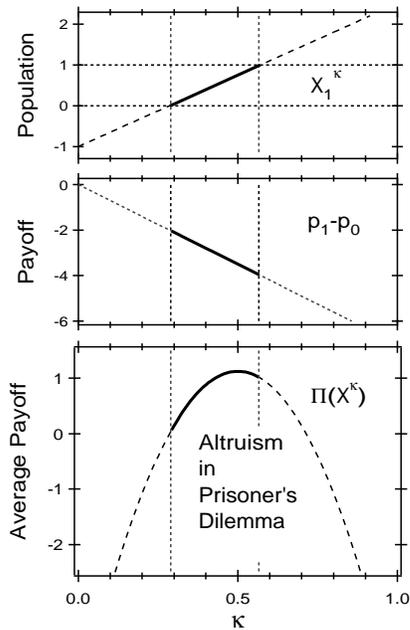}
\caption
{\label{fig1}
The $A^\kappa$-constrained prisoner's dilemma specified by 
the payoff table (13). 
The parameters are chosen to be $\alpha = 4$, $\beta = 1$ and $\gamma = 2$.
In the middle graph, $p_i$ stands for $p_i(A, {\bf X}^\kappa)$.
}
\end{figure}
%
%
%
We illustrate our results with two examples.
First, consider a two-strategy ($i=0, 1$) 
game whose payoffs are
specified by the matrix
%
\begin{eqnarray}
\label{ex1mat}
A = \begin{pmatrix}
      0  & \beta+\alpha \\
      -\gamma & \beta \\
      \end{pmatrix} ,
 \end{eqnarray}
where
$\alpha$, $\beta$ and $\gamma$ are positive real numbers
that satisfy the condition $\alpha$$>\gamma$. 
 This is the famous
example of prisoner's dilemma:  When two players show a
``good hand'', $i=1$, both obtain the payoff of $\beta$, but
when one player betrays the other by playing a bad hand, $i=0$,
he gets the Devil's reward of $\alpha+\beta$ while imposing the
damage $-\gamma$ on the opponent.  When both players show
a ``bad hand'' there is no payoff.
Temporal evolution of $A^\kappa$-constrained game is described by the logistic
equation
%
\begin{eqnarray}
\label{ex1lv}
{\dot x}_1^\kappa  = 
 \left[ \kappa(\alpha+\beta+\gamma)-\gamma\right] x_1^\kappa 
 - (\alpha-\gamma) (x_1^\kappa)^2 .
\end{eqnarray}
The evolutionary Nash equilibrium is  fixed point
$X_1^\kappa$ with an average payoff $\Pi({\bf X}^\kappa)$ 
given respectively by
%
\begin{eqnarray}
\label{ex1x}
X_1^\kappa &=& 
-{ {\gamma} \over {\alpha-\gamma} }
+{ {\alpha+\beta+\gamma} \over {\alpha-\gamma} } \kappa ,
\\ \nonumber
\Pi({\bf X}^\kappa) &=& 
-{ {(\alpha+\beta)\gamma} \over {\alpha-\gamma} }
+{ {(\alpha+\beta+\gamma)^2} \over {\alpha-\gamma} } 
 \kappa(1-\kappa) .
\end{eqnarray}
The per capita system payoff $\Pi({\bf X}^\kappa)$ is indeed 
reflection-symmetric with respect to the line
 $\kappa=1/2$.  
An interesting quantity to look at
is the difference
$p_1(A, {\bf X}^\kappa) - p_0(A, {\bf X}^\kappa)$ 
$=-(\alpha+\beta+\gamma)\kappa$.
This is the payoff disparity between good and bad hands,
which has to be tolerated by good-hand players to achieve
an $A^\kappa$-game with a non-zero value of $\kappa$.  
The peculiarity of this game is that both for a purely egoistic 
game $A^0$ and for a purely altruistic game
$A^1$, the mixed strategy Nash equilibrium (\ref{ex1x}) is 
located outside of a realizable domain $0 \le X_1 \le 1$.  
Easy calculation
shows that only  $\kappa_0 \le \kappa \le \kappa_1$ 
with $\kappa_0 = \gamma/(\alpha+\beta+\gamma)$ and 
$\kappa_1 = \alpha/(\alpha+\beta+\gamma)$ is allowed.
These are the values that 
give $X_1^{\kappa_0}=0$
and $X_1^{\kappa_1}=1$.
If  $\kappa=1/2$ falls between them, the system 
eventually reaches this optimum state having
%
\begin{eqnarray}
\label{ex2pimx}
X_1^{1/2} =
{ {\alpha+\beta-\gamma} \over {2(\alpha-\gamma)} } ,
\quad
\Pi({\bf X}^{1/2}) = { {(\alpha+\beta-\gamma)^2} \over {4(\alpha-\gamma)} } .
\end{eqnarray}
Otherwise, the system settles for $\Pi({\bf X}^{\kappa_1})=\beta$.
Figure 1 depicts an example of the former case.
We expect experimental studies to be performed to check these predictions.
Also, comparison with numerical simulations 
with real strategies with memory (such as Tit-for-Tat or
Pavlov) \cite{AX84,NS93,PO99,KK00} would be beneficial.

As our second example, we consider the following 
three-strategy ($i = 0, 1, 2$) game;
%
\begin{eqnarray}
\label{ex3mat}
A = \begin{pmatrix}
      0  &  0  &  0 \\
      b  & b-a &  b-\rho \\
      -d & f\rho-d & -d \\
      \end{pmatrix} ,
 \end{eqnarray}
where $a$, $b$, $d$, $f$ and $\rho$ are positive real numbers.
With strategy $0$, a player abstains.
With strategy $1$, he/she produces worth valued at $b$. 
Worth is reduced
by $a$ when the opponent also produces worth 
because of overcrowding.
With strategy $2$, the player wastes his/her resources valued at $d$.  
But he/she can derives worth valued at $f\rho$ from 
the worth-producing opponent by way of raiding 
and dimininishing the opponent's worth by $\rho$.
If we describe the system by three-strategy vector ${\bf x}$,
a natural interpretation is
that  $x_1$  represents the portion of total population which
subsists on environmental riches (``commoners''), and $x_2$,
the portion which tries to dominate the opponents (``knights'').
Under the replicator dynamics with
the $A^\kappa$-constrained game, the evolution of the
system is governed by
%
\begin{eqnarray}
\label{ex3lv}
{\dot x}_1^\kappa  &=& 
  (1-\kappa)b x_1^\kappa - a (x_1^\kappa)^2 
  - (1-\kappa-\kappa f)\rho  x_1^\kappa x_2^\kappa 
  \ \ \ \
\\ \nonumber
{\dot x}_2^\kappa &=& 
  -(1-\kappa)d x_2^\kappa 
  + (f-\kappa-\kappa f)\rho x_1^\kappa x_2^\kappa .
\end{eqnarray}
This is the classical Lotka-Volterra prey-predator system, of which
the Nash equilibrium is obtained as
%
\begin{eqnarray}
\label{ex3akap}
X_1^\kappa &=& {{(1-\kappa)d} \over {(f-\kappa-\kappa f)\rho}} ,
\\ \nonumber
X_2^\kappa &=& { {(1-\kappa)}\over{(1-\kappa-\kappa f)\rho} }
  \left[
          b-{ {ad}\over{(f-\kappa-\kappa f)\rho} } 
  \right] .
\end{eqnarray}
One complication is that, when $\kappa$ becomes larger
than $\kappa^*$ $\equiv(f-ad/b\rho)/(1+f)$, 
the fixed point $X_2^\kappa$ falls below zero 
and becomes unstable.  Concurrently, however,
there appears a new trivial Nash equilibrium, which is given by
%
\begin{eqnarray}
\label{ex3akap2}
X_1^\kappa = {{(1-\kappa)b} \over {a}} ,
\quad
X_2^\kappa = 0 .
\end{eqnarray}
This case corresponds to a single-species logistic evolution,
for which the game matrix is effectively reduced to
%
\begin{eqnarray}
\label{ex3mat2}
A^\kappa = \begin{pmatrix}
      0  &  \kappa b   \\
      (1-\kappa)b  & b-a \\
      \end{pmatrix} .
 \end{eqnarray}
In either case, the $A^\kappa$-Nash equilibrium yields 
the per capita system payoff
%
\begin{eqnarray}
\label{ex3pi}
\Pi({\bf X}^\kappa) &=& {D \over T};
\\ \nonumber
D \equiv \det(A^\kappa_{ij}),\ 
&T& \equiv \sum_{i,j}(-1)^{i+j}\Delta_{ij}(A^\kappa) ,
\end{eqnarray}
where $\Delta_{ij}(A^\kappa)$ is the minors of $A^\kappa$.
%
%
\begin{figure}
\includegraphics[width=6.0cm]{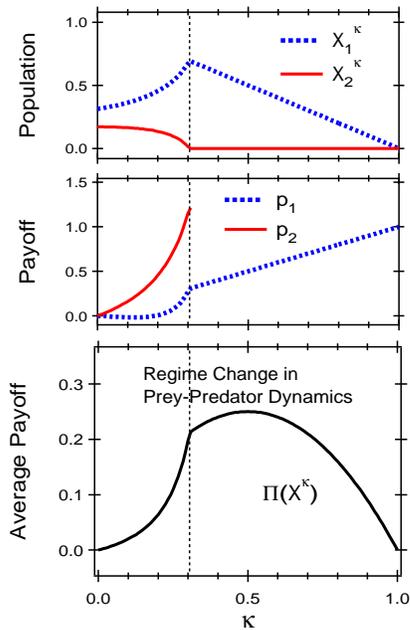}
\caption
{\label{fig2}
The three-strategy game specified by 
the payoff matrix (17) with $A^\kappa$-constrained game dynamics.
The parameters are chosen to be 
$a = 1$, $b = 1$, $d = 1$, $f = 0.8$ and $\rho = 4$.
In the middle graph, 
$p_i$ stands for $p_i(A, {\bf X}^\kappa)$, and $p_0$ is identically zero.
At $\kappa = \kappa^*$, the system displays the transition from
prey-predator dynamics to single spieces logistic dynamics. 
}
\end{figure}

An example of this model with specific parameters is 
depicted in Figure 2.
The $\Pi({\bf X}^\kappa)$ has a single peak at $\kappa=1/2$
as before, but because of the change in the nature of dynamics
at $\kappa^*$, it is no longer symmetric. 
Suppose we start from a knight-commoner dynamics 
of $\kappa=0$ game.
The system is Pareto optimal in that payoffs for both commoners 
and knights are identical. Turning on non-zero $\kappa$ amounts
to introducing ``altruistic culture''.   Curiously, this reduces
the population capacity of knights 
and, at the same time, increase its payoff.
For commoners, it results in lager population and
lower relative payoff.
Although the overall per capita payoff increases quickly, the class
disparity also increases with knights commanding ever higher relative
payoff as the system becomes more altruistic.  
At $\kappa = \kappa^*$ however, the population capacity for knights 
become zero and the ``aristocratic'' regime collapses. 
Above $\kappa^*$,
we have a ``democratic'' regime consisting of single
self-sustaining population of commoners, ever prospering 
with increasing altruism until the system hits the ceiling level
at $\kappa=1/2$, which is the global stability point. 

In both of the above two examples, $\kappa=1/2$ turns out to be 
the only maximum in the allowed region.  But in general, 
higher polynomials in (\ref{ex3pi}) for larger $M$ 
could result in more structures in the
$\Pi({\bf X^\kappa})$ profile for higher numbers of strategies.  
We should thus expect to find more complex
dynamics. 

%
%
%

%
%

%
The indirect payoff maximization through
the ``communal'' arrangement of social goods
is widespread among ecosystems in which components
have intellectual capacity.  
The examples we have studied in this article are
simple toy models which do not necessarily have 
specific real-world counterparts.  
The fact that
they show features 
reminiscent to concepts devised by 
the socio-economic philosophers in past centuries
is rather intriguing.

\medskip
%
%
%
The author wishes to acknowledge his gratitude to Takuma Yamada,
Toshiya Kawai 
and David Greene 
for helpful discussions and useful comments.




\begin{thebibliography}{99}
%
\bibitem{MA74}
R.M. May,
{\em Stability and Complexity in Model Ecosystems},
Priceton Univ. Press, Princeton, 1974).
%
\bibitem{MS82}
J. Maynard Smith,
{\em Evolution and the Theory of Games},
(Cambridge Univ. Press, Cambridge, 1982).
%
\bibitem{GI00}
H. Gintis,
{\em Game Theory Evolving},
(Princeton Univ. Press, Princeton, 2000). 
%
\bibitem{AX84}
R. Axelrod,
{\em The Evolution of Cooperation},
(Basic Books, New York, 1984). 
%
%
\bibitem{BG98}
H. Bester and W. G{\" u}th,
{\em Is altruism evolutionarily stable?}
J. Econ. Behav. Organ. 4 (1998) 193-209.
%
\bibitem{NA50}
J.F. Nash, 
{\em Equilibrium points in n-person games},
Proc. Nat. Acad. Sci. {\bf 36} (1950) 48-49.
%
\bibitem{CH03}
T. Cheon,
{\em Evolutionary stability of ecological hierarchy},
Phys. Rev. Lett.
{\bf 90} (2003) 258105(4).
%
\bibitem{HS88}
J. Hofbauer and K. Sigmund:,
{\em The Theory of Evolution and Dynamical Systems}, 
(Cambridge Univ. Press, Cambridge, 1988).
%
\bibitem{NS93}
M.A. Nowak and K. Sigmund,
{\em A strategy of win stay, lose-shift that outperforms 
tit-for-tat in the prisoner's dilemma}, 
Nature {\bf 364} (1993) 56-58.
%
\bibitem{PO99}
M. Posch, 
{\em Win stay, lose-shift strategies for repeated games -- 
memory length, aspiration levels and noise}, 
J. Theor. Biol. {\bf 198} (1999) 183-195.
%
\bibitem{KK00}
D.P. Kraines and V.Y. Kraines, 
{\em Natural selection of memory-one strategies for 
the iterated prisoner's dilemma}, 
J. Theor. Biol. {\bf 203} (2000) 335-355.
%
%
%
\end{thebibliography}

\end{document}